\documentclass[preprint,superscriptaddress]{revtex4}
\usepackage{graphicx}

\newcommand{\bra}[1] {\left\langle #1 \right|}
\newcommand{\ket}[1] {\left| #1 \right\rangle}

\begin{document}

\title{Fractional squeezing: spectra and dynamics from generalized squeezing Hamiltonian with fractional orders}

\author{Sahel Ashhab}
\affiliation{Advanced ICT Research Institute, National Institute of Information and Communications Technology (NICT), 4-2-1, Nukui-Kitamachi, Koganei, Tokyo 184-8795, Japan}
\affiliation{Research Institute for Science and Technology, Tokyo University of Science, 1-3 Kagurazaka, Shinjuku-ku, Tokyo 162-8601, Japan}

\begin{abstract}
We generalize the generalized-squeezing problem to include fractional values of the squeezing order $n$. This approach allows us to determine the locations of critical points at which qualitative changes in behaviour occur and accurately predict the behaviour at these critical points, which are challenging for conventional computational methods. Based on our numerical calculations, we identify with a high degree of confidence the point at which the spectrum turns from continuous to discrete and the point at which oscillations turn from having asymptotically infinite amplitudes to having finite amplitudes. Furthermore, we numerically investigate the behaviour in the large $n$ regime and provide an intuitive explanation for the numerical results.
\end{abstract}

\maketitle

\newpage

\section{Introduction}
\label{Sec:Introduction}

Driven harmonic oscillators are an essential ingredient in various quantum technologies and are a central subject of study in quantum optics \cite{Walls,Scully}. For example, in superconducting circuit quantum electrodynamics (cQED) systems, electromagnetic modes driven on or near resonance are used for control and readout of qubits, which has made the cQED architecture the most standard one for building superconducting quantum processors \cite{Blais}. In the most basic theoretical model, a photonic mode initially in the vacuum state and driven linearly on resonance will undergo dynamics described by the displacement operator, such that it ends up in a coherent state. Going beyond the linear driving model, parametric driving of photonic modes, in which the mode is driven quadratically at twice its resonance frequency, leads to squeezing, in which photons are generated in pairs \cite{Drummond,Clerk}. The resulting squeezed states have been studied extensively, especially in relation to quantum enhanced sensing. A prime example of the applications of squeezing is its use to increase the sensitivity of gravitational wave detectors \cite{Jia}. Another emerging application of squeezed states is quantum error correction \cite{Gottesman,LachanceQuirion,Brock}. The maximum squeezing achieved to date is 15 dB \cite{Vahlbruch}. Generalized squeezing, in which photons are generated in multiples of $n$ photons, is gaining increasing attention in recent years, spurred in part by the rapid recent advances in developing higher-order nonlinear interactions in various quantum technology platforms, such as superconducting circuits \cite{Chang,Eriksson,Felicetti,Ayyash2024} and trapped ions \cite{Bazavan,Saner}. While higher-order squeezing was demonstrated in recent experiments, as evidenced by the observation of the characteristic three- and four-arm phase-space functions, these experimental demonstrations remain at the level of a few photons \cite{Eriksson,Saner}. Therefore, much more work remains to be done on the experimental side before this technology reaches the level of practical applications.

Furthermore, while generalized squeezing might at first sight look like a straightforward generalization of two-photon squeezing, it turns it to be much more difficult to model and analyze theoretically than two-photon squeezing \cite{Fisher,Hillery1984,Hong,Braunstein1987,Braunstein1990,Hillery1990,Braak}. These difficulties are highlighted by the contradicting conclusions of recent studies on the subject \cite{Ashhab2025,Gordillo,Ashhab2026PRA,Fischer,Ashhab2026NJP}. In contrast, displacement and two-photon squeezing allow the derivation of exact analytical formulae that describe how quantum states are transformed under these operations \cite{Walls,Scully}. 

One of the interesting aspects of generalized squeezing observed in the simulations of Ref.~\cite{Ashhab2025} is the qualitative differences in behaviour for different squeezing orders. Some of the changes are counter-intuitive. For example, while the number of photons grows indefinitely and approaches infinity in the infinite-time limit for two-photon squeezing ($n=2$), the photon number diverges at a finite time for tri-squeezing ($n=3$), and the photon number remains finite at all times for quint-squeezing ($n=5$). Such disparate behaviour makes it highly desirable to have a better understanding of how the evolution between different squeezing orders happens.

While the squeezing order is an integer, we show in this paper that calculations can be set up in which the squeezing order can take any real-number value in a range that covers all integers from 1 to infinity. These calculations give us a closer view of how the behaviour changes as we change the squeezing order. Furthermore, it provides more certainty about the behaviour in certain cases than what we would obtain by analyzing integer values only. One example is the continuity of the spectrum in the case of two-photon squeezing. The simulations in Ref.~\cite{Ashhab2026PRA} were inconclusive, because the convergence was so slow that it was not possible to reliably extrapolate the simulation results and infer the asymptotic values of the squeezing Hamiltonian's eigenvalues. In this work, by analyzing the overall landscape of the eigenvalues as functions of simulation size and squeezing order, we obtain much stronger evidence that the two-photon-squeezing spectrum is continuous, which is consistent with the well-known behaviour of two-photon squeezing dynamics. A similar situation arises in the case of four-photon squeezing: extrapolating simulation results is challenging when the squeezing order is exactly equal to four, but the overall landscape becomes much simpler when we allow the squeezing order to be an additional variable parameter and we consider values that are near but not exactly equal to four. We will show that the cases of two- and four-photon squeezing represent critical points in the behaviour of some physical quantities, which explains why these two cases are especially challenging for numerical simulations.

The remainder of this paper is organized as follows: In Sec.~\ref{Sec:Theory}, we describe the theoretical background and methodology of our calculations. In Sec.~\ref{Sec:Simulations}, we present the results of our numerical simulations and discuss some of their implications, such as the boundary between continuous and discrete spectra and the amplitude of generalized-squeezing oscillations. We discuss a few additional aspects of the results in Sec.~\ref{Sec:Discussion}, such as the asymptotic behaviour at large values of the squeezing order. We give some final remarks in Sec.~\ref{Sec:Conclusion}.

\section{Theoretical background and computational methodology}
\label{Sec:Theory}

The generalized squeezing operator with squeezing order $n$ is given by
\begin{equation}
\hat{U}_n \left( r \right) = \exp \left\{ -i r \hat{H}_n \right\},
\label{Eq:GeneralizedSqueezingOp}
\end{equation}
where the squeezing Hamiltonian is given by
\begin{equation}
\hat{H}_n = i \left[ \left(\hat{a}^\dagger\right)^n - \hat{a}^n \right],
\label{Eq:Hamiltonian_n}
\end{equation}
$r$ is the squeezing parameter, and $\hat{a}$ and $\hat{a}^{\dagger}$ are, respectively, the photon annihilation and creation operators.

As discussed in Ref.~\cite{Fischer}, the squeezing Hamiltonian is generally not essentially self-adjoint, and extra care must be taken when performing various operations on it to predict the behaviour of a physical system undergoing generalized squeezing dynamics. In this work, we will focus on truncated versions of the operator and therefore not worry about the subtle mathematical questions related to self-adjointness, since these questions are not related to the main purpose of this work. Furthermore, we will focus on the squeezed vacuum state $\ket{\psi_n(r)}$, which is obtained by applying the squeezing operator to the vacuum (i.e.~zero-photon) state $\ket{0}$:
\begin{equation}
\ket{\psi_n(r)} = \hat{U}_n \left( r \right) \ket{0}.
\end{equation}

The relevant basis that contains the vacuum state is $\left\{ \ket{0}, \ket{n}, \ket{2n}, \ldots \right\}$. The Hamiltonian for this space can be expressed in matrix form as
\begin{equation}
\hat{H}_n = \left(
\begin{array}{ccccc}
0 & -i \sqrt{n!} & 0 & 0 & \cdots \\
i \sqrt{n!} & 0 & -i \sqrt{\frac{(2n)!}{n!}} & 0 & \\
0 & i \sqrt{\frac{(2n)!}{n!}} & 0 & -i \sqrt{\frac{(3n)!}{(2n)!}} & \\
0 & 0 & i \sqrt{\frac{(3n)!}{(2n)!}} & 0 & \\
\vdots & & & & \ddots
\end{array}
\right).
\label{Eq:Hamiltonian_n_Matrix}
\end{equation}
As mentioned above, we would like to analyze the evolution of various quantities as we gradually vary the squeezing order $n$. For this purpose, we rewrite Eq.~(\ref{Eq:Hamiltonian_n_Matrix}) in the equivalent alternative form
\begin{equation}
\hat{H}(n) = \left(
\begin{array}{ccccc}
0 & -i \sqrt{\frac{\Gamma(n+1)}{\Gamma(1)}} & 0 & 0 & \cdots \\
i \sqrt{\frac{\Gamma(n+1)}{\Gamma(1)}} & 0 & -i \sqrt{\frac{\Gamma(2n+1)}{\Gamma(n+1)}} & 0 & \\
0 & i \sqrt{\frac{\Gamma(2n+1)}{\Gamma(n+1)}} & 0 & -i \sqrt{\frac{\Gamma(3n+1)}{\Gamma(2n+1)}} & \\
0 & 0 & i \sqrt{\frac{\Gamma(3n+1)}{\Gamma(2n+1)}} & 0 & \\
\vdots & & & & \ddots
\end{array}
\right).
\label{Eq:Hamiltonian_n_MatrixGamma}
\end{equation}
While the factorial function is defined only for non-negative integers, the gamma function is defined for all real numbers. We can therefore treat $n$ as a continuous variable that can take any positive value (or, in principle, even negative values for which the gamma function is finite). In other words, the squeezing order $n$ is no longer confined to integer values.

There is a serious conceptual issue with allowing $n$ to take fractional values. When $n$ is an integer, we can intuitively understand the Hamiltonian matrix as describing processes in which $n$ photons are created or annihilated. When $n$ is a fractional number, there is no simple, intuitive way to describe the Hamiltonian matrix, or even the basis, in physical terms. We therefore think of fractional $n$ values as a mathematical tool that allows us to analyze the changes in various quantities as we gradually vary the squeezing order $n$.

We should mention here the field of fractional calculus, in which one can take fractional derivatives, as opposed to simply taking the first derivative, second derivative, etc. \cite{Laskin,Herrmann}. There could be methods and results in the field of fractional calculus that can be applied to the present problem. In this context, one could write the operators in the real-space representation [e.g., $\hat{a}=\left(x + \partial/\partial x \right)/\sqrt{2}$] and obtain the Schr\"odinger equation
\begin{equation}
i \frac{\partial\psi(x,t)}{\partial t} = \frac{i}{2^{n/2}} \left[ \left(x - \frac{\partial}{\partial x} \right)^n - \left(x + \frac{\partial}{\partial x} \right)^n \right] \psi(x,t).
\label{Eq:SchroedingerEquationRealSpace}
\end{equation}
When the squeezing order $n$ is a fractional number, we obtain derivatives raised to partial powers. Investigating these fractional derivatives is the subject of fractional calculus. This endeavour can lead to involved mathematical techniques that go beyond our knowledge about the subject. Considering that exploring the fractional calculus aspect of the problem is not necessary for establishing the results that we are seeking in the present work, we will not pursue this issue any further. It is also worth mentioning here the somewhat related idea of treating the number of spatial dimensions as a variable parameter that can take any value, including fractional ones \cite{Goldenfeld}. Some calculations are then simplified by first treating the infinite-dimension limit and then using the results as a starting point to investigate three-dimensional systems.

Our simulations proceed as follows: We construct a truncated Hamiltonian matrix of size $N \times N$. We always choose even values of $N$. To obtain the spectrum, we diagonalize this truncated Hamiltonian. As discussed in Ref.~\cite{Ashhab2026PRA}, the two eigenvalues that are closest to zero (and located symmetrically on the positive and negative sides of zero) contain important information, for example whether the spectrum is continuous or discrete in the limit of infinite $N$. For questions related to the dynamics, considering the fact that the vacuum state is typically well approximated by a superposition of only the two eigenstates of the Hamiltonian that correspond to the above-mentioned eigenvalues, we analyze these eigenstates. For the purpose of analyzing the dependence on $n$, we perform simulations for $n=0, 0.1, 0.2, \ldots, 6$.

When analyzing the quantum dynamics of a photonic state, one of the standard physical quantities to analyze is the photon number operator $\hat{a}^{\dagger} \hat{a}$. However, this operator is associated with some conceptual complications. The most obvious complication occurs when $n=0$. In this case, the basis states are all separated by zero photons, and the off-diagonal matrix elements in the Hamiltonian create or annihilate zero photons. In other words, the number of photons is always equal to zero, regardless of the quantum state. For this reason, we define the renormalized number operator
\begin{equation}
\hat{m} = \left(
\begin{array}{ccccc}
0 & \;\; 0 & \;\; 0 & \;\; 0 & \cdots \\
0 & \;\; 1 & \;\; 0 & \;\; 0 & \\
0 & \;\; 0 & \;\; 2 & \;\; 0 & \\
0 & \;\; 0 & \;\; 0 & \;\; 3 & \\
\vdots & & & & \ddots \\
\end{array}
\right).
\end{equation}
For (nonzero) integer squeezing orders $n$, the usual photon number operator $\hat{a}^{\dagger} \hat{a}$ is related to $\hat{m}$ by the simple formula $\hat{a}^{\dagger} \hat{a} = n \hat{m}$, such that both operators essentially convey the same information, and they are both physically meaningful. When $n=0$, analyzing the operator $\hat{m}$ obviously provides more information than analyzing the operator $n\hat{m}$, which is always equal to zero. The expectation value $\left\langle \hat{m} \right\rangle = \bra{\psi_n^{(N)}(r)} \hat{m} \ket{\psi_n^{(N)}(r)}$ will be one of the main quantities that we will analyze below. We will also show one plot for the quantity $n\left\langle \hat{m} \right\rangle$, which we will also express as $\left\langle \hat{a}^{\dagger} \hat{a} \right\rangle$, to provide an intuitively clear presentation of the results.

\section{Simulation results}
\label{Sec:Simulations}

We now present the results of the simulations described in Sec.~\ref{Sec:Theory}.

\subsection{Continuity of spectrum}
\label{Sec:SpectrumContinuity}

\begin{figure}[h]
\includegraphics[width=8cm]{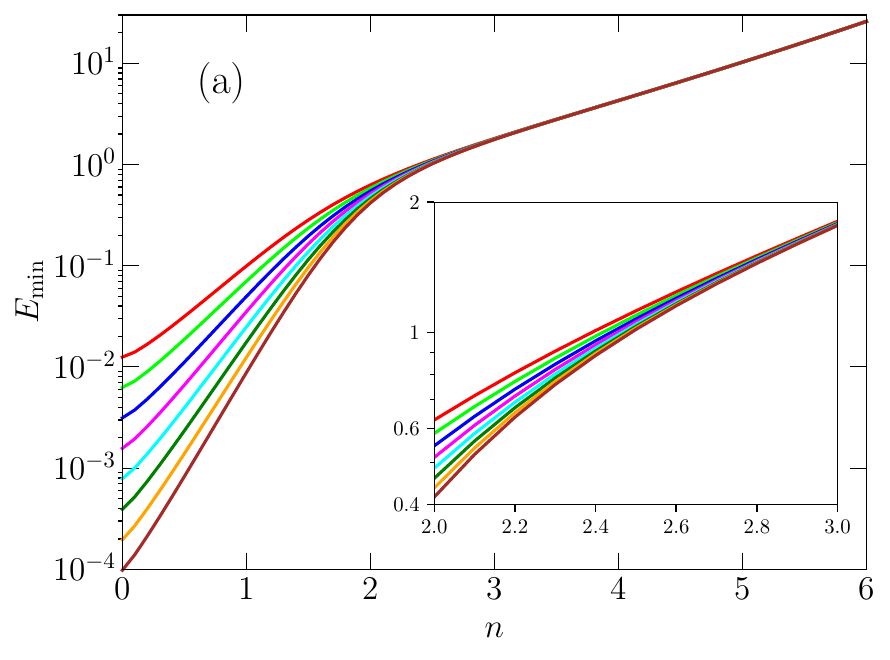}
\includegraphics[width=8cm]{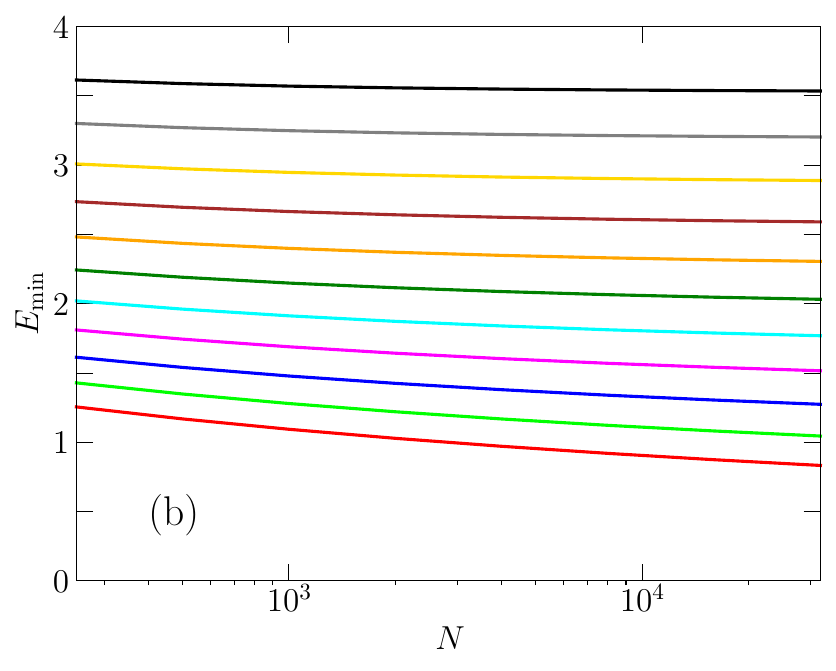}
\caption{(a) Smallest positive eigenvalue of the Hamiltonian $\hat{H}(n)$, which we denote as $E_{\rm min}$, as a function of the squeezing order $n$. The different lines (from top to bottom, i.e.~red, green, blue, ...) correspond to to truncation sizes $N=250, 500, 1000, ..., 250 \times 2^7$. The inset zooms in on the range $2 \leq n \leq 3$. The progression of these lines shows that for small values of $n$ (visually, $n\lesssim 2$), the eigenvalue has not converged yet and appears to have the asymptotic values zero, while for large values of $n$ ($n\gtrsim 3$) the eigenvalue has clearly converged to its asymptotic value. As a further visual depiction of the convergence behaviour, Panel (b) shows $E_{\rm min}$ as a function of the Hamiltonian truncation size $N$ for $n=2, 2.1, 2.2,..., 3$ (bottom to top, i.e.~red, green, blue, ...).}
\label{Fig:EigenvalueVsSqueezingOrderAndTruncationSize}
\end{figure}

\begin{figure}[h]
\includegraphics[width=8cm]{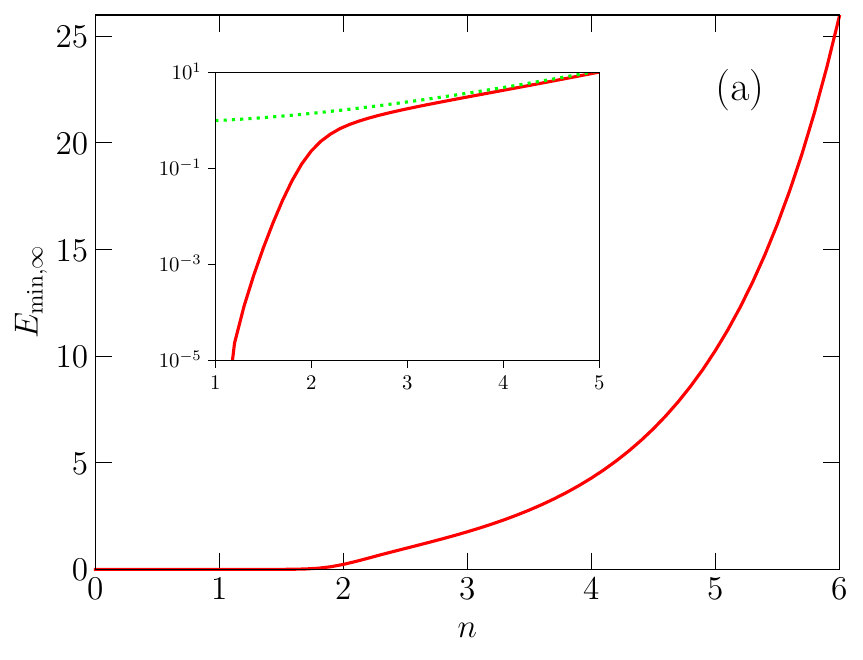}
\includegraphics[width=8cm]{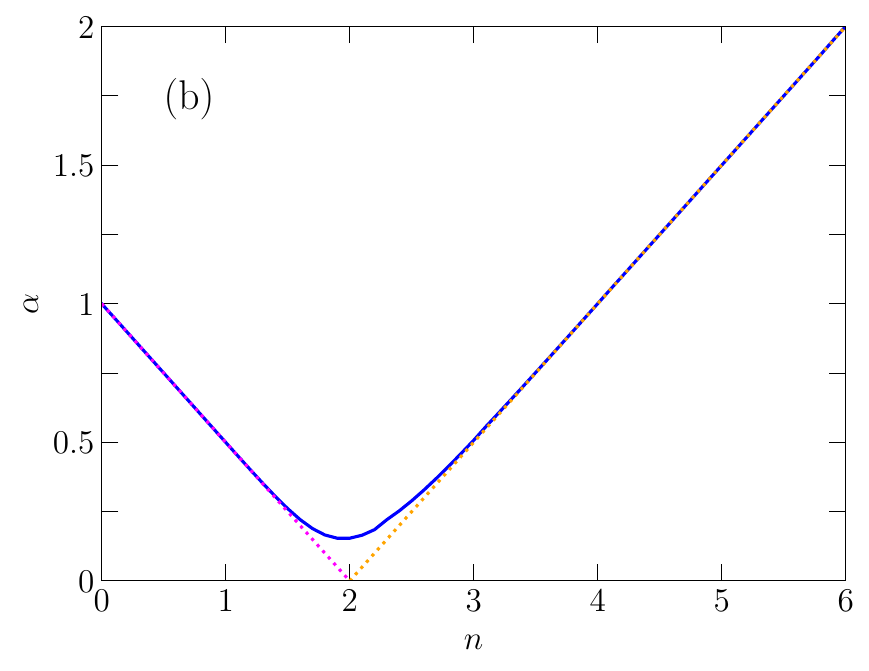}
\caption{(a) The asymptotic value of $E_{\rm min}$, obtained by extrapolating the data in Fig.~\ref{Fig:EigenvalueVsSqueezingOrderAndTruncationSize} to $N\to\infty$ and denoted by $E_{\rm min,\infty}$, as a function of the squeezing order $n$. Starting from $n=0$, $E_{\rm min,\infty}$ remains essentially at zero up to around $n=2$, after which it rises rapidly. The dependence of $E_{\rm min,\infty}$ on $n$ is examined further in the inset, which uses a logarithmic scale for the $y$ axis. The green line is given by $\sqrt{\Gamma(n+1)}$, i.e.~the smallest nonzero matrix element in the Hamiltonian. We will explain in Sec.~\ref{Sec:Discussion} why this function gives the asymptotic value of $E_{\rm min,\infty}$ in the large-$n$ limit. (b) The exponent $\alpha$ obtained from fitting the $(N,E_{\rm min})$ data to the function in Eq.~(\ref{Eq:EminFittingFunction}). Away from $n=2$, $\alpha$ clearly follows linear functions of $n$, one for $n<2$ and one for $n>2$. The linear dependence is represented by the dotted lines. As we approach $n=2$, we see deviations from the linear functions. This deviation is likely caused by the inevitable numerical errors in our finite numerical simulations. Both linear functions go through the point $(n=2,\alpha=0)$ when extrapolated.}
\label{Fig:EigenvalueFittingParametersVsSqueezingOrder}
\end{figure}

We start with the question of whether the spectrum is continuous or discrete. In Fig.~\ref{Fig:EigenvalueVsSqueezingOrderAndTruncationSize}(a), we plot the smallest positive eigenvalue of the Hamiltonian $\hat{H}(n)$, which we denote as $E_{\rm min}$, as a function of the squeezing order $n$. At the extreme point $n=0$, it is clear that $E_{\rm min}$ decreases by the same factor every time we double $N$, which implies that $E_{\rm min}\propto N^{-\alpha}$ with some exponent $\alpha$. As such, the asymptotic value of $E_{\rm min}$ in the limit $N\to\infty$ is zero, and the spectrum is continuous in this limit. At the opposite extreme ($n=6$), the spectrum barely changes as we increase $N$ from 250 to $250 \times 10^7$, indicating that the spectrum is discrete. In the intermediate region, roughly from $n=2$ to $n=3$, the behaviour is not completely clear from a visual inspection of Fig.~\ref{Fig:EigenvalueVsSqueezingOrderAndTruncationSize} alone. To have a better visual view of the convergence behaviour in the region $2\leq n \leq 3$, we plot $E_{\rm min}$ as a function of $N$ in Fig.~\ref{Fig:EigenvalueVsSqueezingOrderAndTruncationSize}(b). We can see that there is a trend towards a larger asymptotic value as we increase $n$, but we cannot make a definite conclusion about the asymptotic value for each curve, especially near $n=2$.

Next we perform a fitting of the $(N,E_{\rm min})$ data (with a fixed value of $n$ for each data set) to the function
\begin{equation}
E_{\rm min} = E_{\rm min,\infty} + \delta E_{\rm min} \times N^{-\alpha}.
\label{Eq:EminFittingFunction}
\end{equation}
The constant term ($E_{\rm min,\infty}$) is the asymptotic value of $E_{\rm min}$ in the limit $N\to\infty$, while the exponent $\alpha$ quantifies how fast $E_{\rm min}$ approaches its asymptotic value. The results are shown in Fig.~\ref{Fig:EigenvalueFittingParametersVsSqueezingOrder}. The asymptotic value $E_{\rm min,\infty}$ is zero from $n=0$ all the way up to slightly below $n=2$. For $n>2$, $E_{\rm min,\infty}$ increases rapidly and approaches the function $\sqrt{\Gamma(n+1)}$, which is the smallest nonzero matrix element in the Hamiltonian. We will discuss in Sec.~\ref{Sec:Discussion} the reason why this function describes this asymptotic behaviour of $E_{\rm min,\infty}$. It is worth noting that the large-$n$ limit of this function is given by Stirling's formula:
\begin{equation}
\Gamma(n+1) \sim \sqrt{2\pi n} \left( \frac{n}{e} \right)^n,
\end{equation}
which increases super-exponentially as a function of $n$. The exponent $\alpha$ starts off at $\alpha=1$ when $n=0$, then decreases linearly as a function of $n$ in the direction of the point $(n=2,\alpha=0)$. Slightly before reaching $n=2$, the line bends and does not reach exactly $\alpha=0$. It is worth noting here that $\alpha=0$ means that $E_{\rm min}$ is independent of $N$. However, this situation could be equally well fitted with a function that has $\delta E_{\rm min}=0$. In other words, this situation would create complications for the fitting procedure. We do not need to worry about this situation, because we did not encounter it in the data. When $n$ is sufficiently larger than 2, $\alpha$ follows a linearly increasing function of $n$. Upon extrapolation, both linear functions that accurately describe $\alpha$ well below and well above $n=2$ predict that $\alpha=0$ at $n=2$. Taking the two panels of Fig.~\ref{Fig:EigenvalueFittingParametersVsSqueezingOrder} together, we can infer that $E_{\rm min,\infty}$ is likely to be equal to zero for all values of $n$ between $n=0$ and $n=2$, with a sub-polynomial scaling law governing $E_{\rm min}$ as a function of $N$ exactly at $n=2$. The deviation of our results from this scenario can be attributed to numerical errors in our finite simulations.

\subsection{Quantum state size}
\label{Sec:EigenstateSize}

\begin{figure}[h]
\includegraphics[width=8cm]{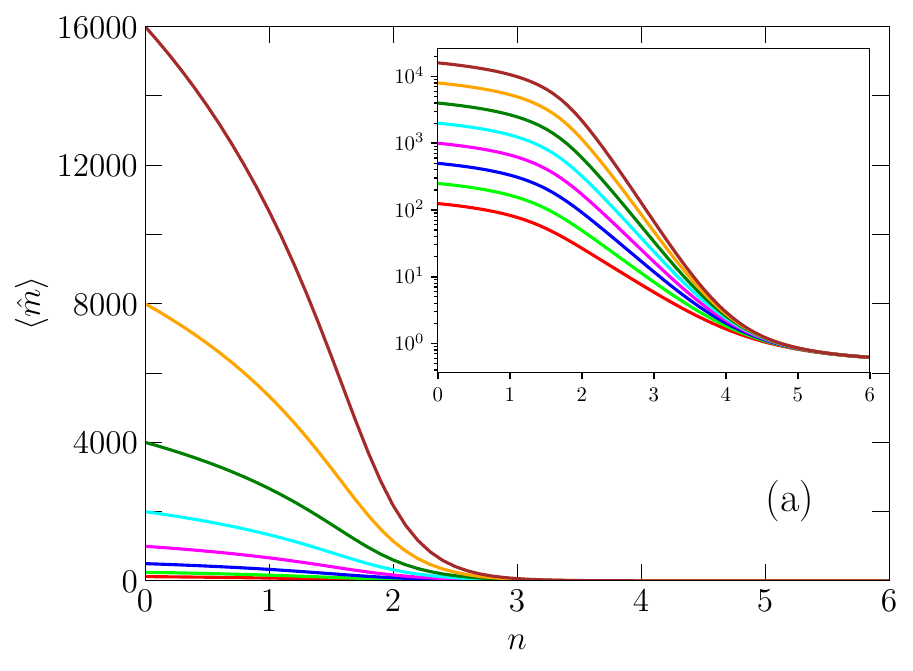}
\includegraphics[width=8cm]{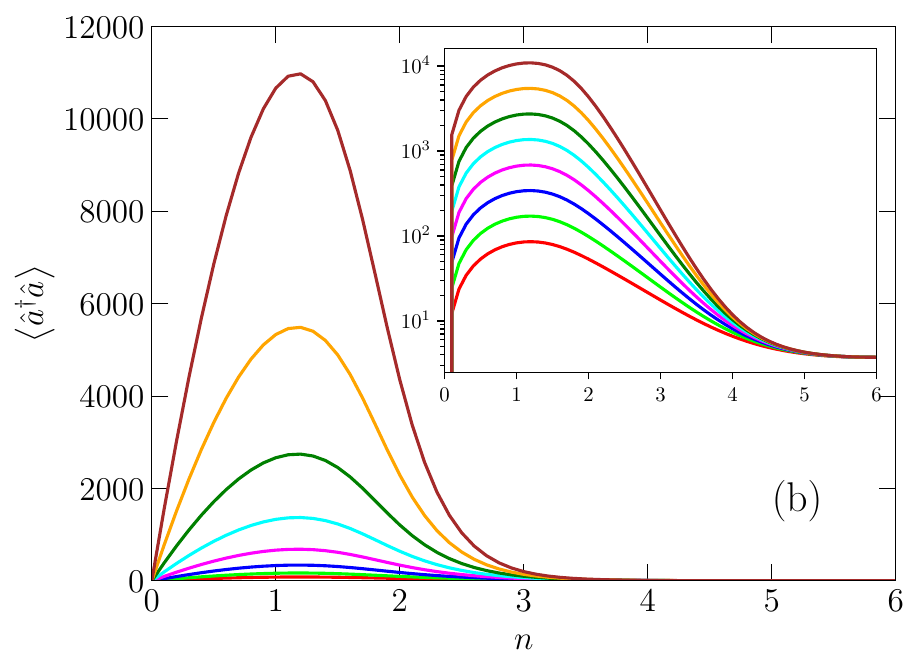}
\caption{(a) Expectation value of the renormalized photon number operator $\left\langle \hat{m} \right\rangle$ for the eigenstate that corresponds to the eigenvalue $E_{\rm min}$ as a function of squeezing order $n$. Panel (b) is a similar plot for the operator $n \hat{m}$, which we express as the usual photon number operator $\hat{a}^{\dagger} \hat{a}$. Both panels show that the plotted quantities seem to increase indefinitely as functions of $N$ when $n<4$. When $n>5$, there is no discernible dependence on $N$ for our simulation data, in which $N$ covers the range $[250,250\times 2^7]$. The functional dependence is analyzed in more detail in Figs.~\ref{Fig:EigenstateSizeVsTruncationSize} and \ref{Fig:EigenstateSizeFittingParametersVsSqueezingOrder}.}
\label{Fig:EigenstateSizeVsSqueezingOrder}
\end{figure}

\begin{figure}[h]
\includegraphics[width=8cm]{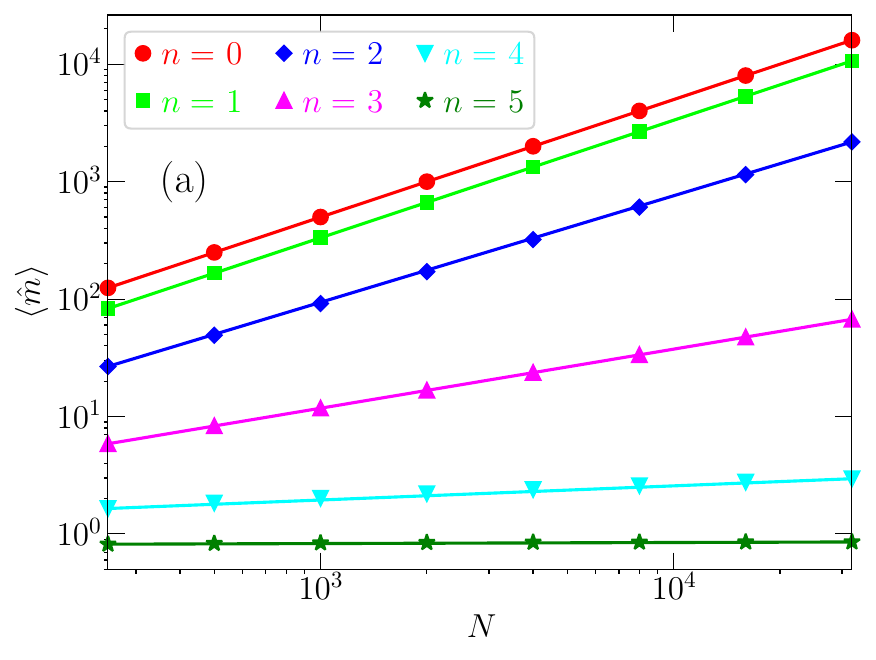}
\includegraphics[width=8cm]{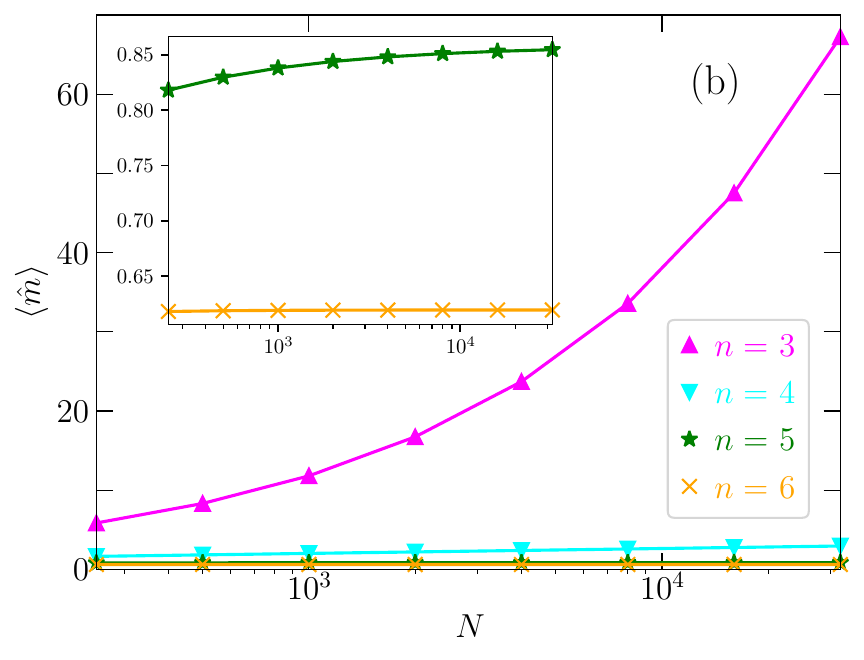}
\caption{$\left\langle \hat{m} \right\rangle$ as a function of the truncation size $N$ for a few values of $n$. Panel (a) shows the data in a log-log plot, while Panel (b) shows the data in a log-linear plot. Both panels show that $\left\langle \hat{m} \right\rangle$ increases indefinitely as a function of $N$ up to $n=4$, while it converges to finite values when $n=5,6$. The fact that each data set up to $n=3$ follows a straight line in the log-log plot indicates that the data follows a power law. The case $n=4$ was analyzed in Ref.~\cite{Ashhab2025}, where it was found that the scaling is logarithmic. The logarithmic scaling cannot be seen clearly in this figure. The inset in Panel (b) shows that $\left\langle \hat{m} \right\rangle$ converges to finite values for $n=5,6$. This point will be shown more clearly in Fig.~\ref{Fig:EigenstateSizeFittingParametersVsSqueezingOrder}.}
\label{Fig:EigenstateSizeVsTruncationSize}
\end{figure}

\begin{figure}[h]
\includegraphics[width=8cm]{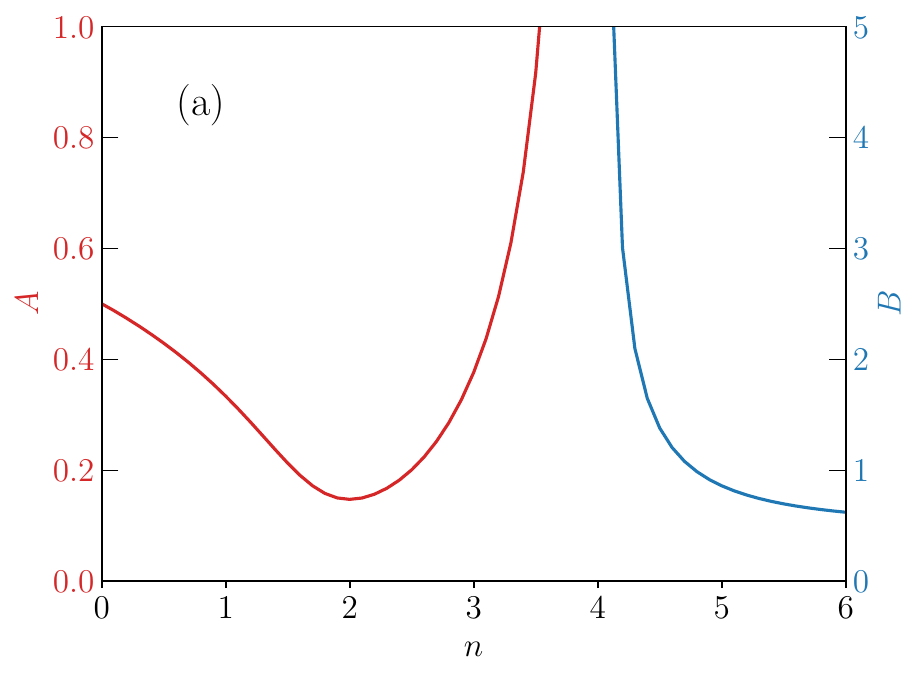}
\includegraphics[width=8cm]{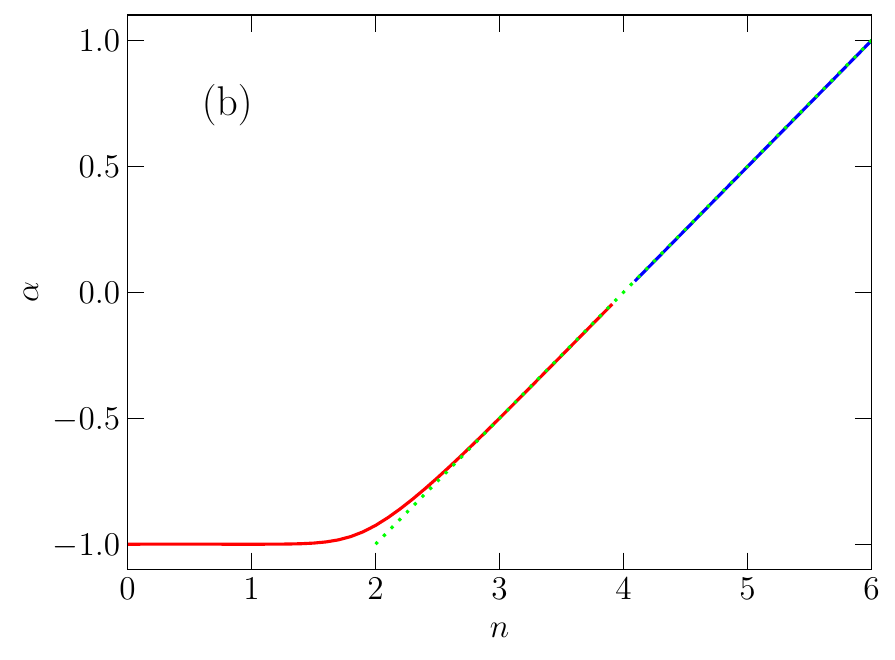}
\caption{Fitting parameters obtained by fitting the $(N,\left\langle \hat{m} \right\rangle)$ data to the function $A\times N^{-\alpha}+B$. As can be seen in Panel (b), the exponent $\alpha$ remains essentially constant from $n=0$ to $n=2$. Then it increases linearly. The line interruption and change in color at $n=4$ signal that the point $n=4$ represents a demarcation point. When $n<4$, $\alpha$ is negative, which means that $\left\langle \hat{m} \right\rangle$ increases indefinitely and has the asymptotic value infinity. When $n>4$, $\alpha$ is positive, which means that $\left\langle \hat{m} \right\rangle$ approaches a finite asymptotic value. For this reason, it makes sense to focus on the parameter $A$ (i.e.~the scaling prefactor) for $n<4$ and to focus on the parameter $B$ (i.e.~the asymptotic value) for $n>4$, as shown in Panel (a). The parameter $B$ approaches 0.5 in the large-$n$ limit, as will be discussed in Sec.~\ref{Sec:Discussion}. The dotted green line in Panel (b) is a straight line between the points $(2,-1)$ and $(6,1)$. It serves as a guide to the eye, to show how well the red and blue lines follow a straight line.}
\label{Fig:EigenstateSizeFittingParametersVsSqueezingOrder}
\end{figure}

We now analyze the expectation value of the renormalized photon number operator $\left\langle \hat{m} \right\rangle$ for the eigenstate that corresponds to the smallest positive eigenvalue, i.e.~$E_{\rm min}$. This quantity describes the size, or spread, of the quantum state. It is also closely related to the oscillation amplitude in the dynamics for $n>2$, where the squeezing dynamics exhibits oscillations \cite{Ashhab2025}. Specifically, the photon number oscillation amplitude is $2n\left\langle \hat{m} \right\rangle$, up to corrections from smaller, fast-oscillating terms.

Figure \ref{Fig:EigenstateSizeVsSqueezingOrder} shows that $\left\langle \hat{m} \right\rangle$ decreases monotonically as a function of $n$ for any given value of $N$. At the extreme point $n=0$, $\left\langle \hat{m} \right\rangle = N/2$. This result can be intuitively understood by noting the fact that all the nonzero matrix elements in the Hamiltonian $\hat{H}(0)$ are equal, and there is a mirror symmetry between low and high number states. If we fix $n$ somewhere in the range $0 \leq n \leq 3$ and increase $N$, we see that $\left\langle \hat{m} \right\rangle$ increases indefinitely. This behaviour is seen more clearly in Fig.~\ref{Fig:EigenstateSizeVsTruncationSize}. The data in Fig.~\ref{Fig:EigenstateSizeVsTruncationSize}, as well as all other $(N,\left\langle \hat{m} \right\rangle)$ data sets extracted from Fig.~\ref{Fig:EigenstateSizeVsSqueezingOrder}, are fitted well by the function
\begin{equation}
\left\langle \hat{m} \right\rangle = A \times N^{-\alpha} + B.
\end{equation}
The exponent $-\alpha$, shown in Fig.~\ref{Fig:EigenstateSizeFittingParametersVsSqueezingOrder}, exhibits three different types of behaviour, which divides the set of all $n$ values into three regimes. This exponent remains constant ($-\alpha=1$) from $n=0$ to $n=2$. This result means that the eigenstate size is linearly proportional to $N$ in this regime, even though the overall scale (given by the parameter $A$) decreases with increasing $n$. Between $n=2$ and $n=4$, $\alpha$ increases linearly towards $\alpha=0$ at $n=4$. This trend indicates that the scaling law of $\left\langle \hat{m} \right\rangle$ as a function of $N$ slows down as $n$ increases. Exactly at $n=4$, the fitting fails and generates an error in our numerical fitting calculations. The reason is that the data follows a logarithmic scaling law, as shown in Ref.~\cite{Ashhab2025}, and here we try to fit it with a power-law function. When $n>4$, the $(N,\left\langle \hat{m} \right\rangle)$ data have finite asymptotic values, which translates into positive values of $\alpha$. It is interesting that the straight lines below and above $n=4$ follow the same function, even though there is a qualitative change in behaviour at $n=4$. As mentioned above, the scaling prefactor $A$ [Fig.~\ref{Fig:EigenstateSizeFittingParametersVsSqueezingOrder}(a)] decreases from $n=0$ to $n=2$ and then increases from $n=2$ to $n=4$. It should be noted that for $n\leq 2$ the squeezing dynamics is expected to exhibit a monotonic increase in the photon number, while the dynamics is expected to exhibit oscillations for $n>2$, which would complicate any comparison in the behaviour of the parameter $A$ across the two regions. Furthermore, for $n<2$, the vacuum state has non-negligible overlap with a large number of eigenstates, such that the properties of the two eigenstates at the center of the spectrum do not provide a sufficient description of the dynamics. Note also that the increase in $A$ from $n=2$ to $n=4$ is accompanied by an increase in $\alpha$, such that $\left\langle \hat{m} \right\rangle$ and $\left\langle \hat{a}^{\dagger} \hat{a} \right\rangle$ end up being by monotonically decreasing functions of $n$ for sufficiently large values of $N$, as can be seen in Fig.~\ref{Fig:EigenstateSizeVsSqueezingOrder}. The asymptotic value of $\left\langle \hat{m} \right\rangle$, i.e.~the parameter $B$ in Fig.~\ref{Fig:EigenstateSizeFittingParametersVsSqueezingOrder}, approaches the value 0.5 for large $n$. The interpretation of this result will be discussed in Sec.~\ref{Sec:Discussion}.

\section{Discussion}
\label{Sec:Discussion}

In this section we discuss a few additional aspects of our results.

\subsection{Physical mechanisms of transitions at $n=2$ and $n=4$}

Our results in Sec.~\ref{Sec:Simulations} revealed two critical points at which sudden transitions in the physical behaviour occur. These points are $n=2$ and $n=4$. The physical mechanisms of the transitions at these two points are qualitatively different. At $n=2$, both the eigenvalues at the middle of the spectrum and the sizes of the corresponding eigenstates exhibit sudden changes in their behaviour. The spectrum changes from being continuous to being discrete. One could say that this transition is the physically more significant one, because it indicates a transition from monotonically increasing photon number to oscillatory dynamics. At $n=4$, the spectrum, or at least the eigenvalues at the center of the spectrum, do not show any special features when $n$ is varied across $n=4$. In spite of the smooth behaviour of the eigenvalues, the size of the eigenstates exhibits a drastic transition from being divergent as a function of $N$ to being finite. It is indeed surprising that this dramatic transition is not reflected at all in the eigenvalues. It should also be emphasized that the finite asymptotic values of the eigenstate size when $n>4$ do not mean that they become independent of $N$ in the limit $N\to\infty$. They still depend on whether $N$ is even or odd, as was discussed in detail in Ref.~\cite{Ashhab2026PRA}.

\subsection{Asymptotic behaviour in large-$n$ limit}

The results presented in Sec.~\ref{Sec:Simulations} indicate a trend towards simple asymptotic behaviour in the large-$n$ limit. Specifically, $E_{\rm min,\infty}$ approaches $\sqrt{\Gamma(n+1)}$ and the parameter $B$ in Fig.~\ref{Fig:EigenstateSizeFittingParametersVsSqueezingOrder} approaches 0.5. These results are what we would obtain if we truncate the Hamiltonian at $N=2$ and ignore all basis states except the first two. An intuitive understanding of this situation can be achieved by considering the artificial but instructive Hamiltonian
\begin{equation}
\hat{H}_{\rm hierarchical} = \left(
\begin{array}{cccccc}
0 & h_2 & 0 & \cdots & 0 & 0 \\
h_2 & 0 & h_3 & & 0 & 0 \\
0 & h_3 & 0 & & 0 & 0 \\
\vdots & & & \ddots & & \vdots \\
0 & 0 & 0 &  & 0 & h_N \\
0 & 0 & 0 & \cdots & h_N & 0 \\
\end{array}
\right),
\end{equation}
where $h_j$ are all real and positive, and $h_2 \ll h_3 \ll \cdots \ll h_N$. If we focus on the highest two Fock states, they are coupled in the Hamiltonian by the pair of matrix elements equal to $h_N$. As a result, the Hamiltonian has a pair of large (positive and negative) eigenvalues approximately given by $\pm h_N$. The corresponding eigenvectors are, to a good approximation, localized at the two highest states in the Fock space. All other matrix elements in the Hamiltonian are much smaller than $h_N$. Because of the energy mismatch, the two states with energies $\pm h_N$ are almost completely decoupled from the rest of the Hilbert space. We can therefore remove them from the Hamiltonian, if we are considering dynamics and spectra that do not reach the energy scale of $h_N$. Furthermore, considering that all eigenvectors must be orthogonal to each other, the $(N-2)$ remaining eigenvectors must have a small total weight in the highest two Fock states. We can therefore restrict ourselves to the state space extending from 1 to $N-2$, as the two highest states will not play a role near the middle of the spectrum. We can then repeat the same procedure and gradually remove pairs of eigenvalues from the spectrum until we are left with one or two eigenvalues at the center of the spectrum, depending on whether $N$ is odd or even. The last remaining eigenvectors must be mostly confined to the lowest one or two Fock states. We have performed numerical simulations with matrix sizes ranging from $2 \times 2$ to $8 \times 8$ and with $h_{j+1}/h_j=10$ or $100$ and confirmed that the spectrum follows the pattern explained above, i.e.~forming energy level pairs with energies $\pm h_N$, $\pm h_{N-2}$, $\pm h_{N-4}$ etc. When $N$ is even, the last two eigenvalues in the middle of the spectrum will be given by $\pm h_2$, and the corresponding eigenvectors will be equal superpositions of the two lowest Fock states. When $N$ is odd, the last remaining eigenvalue is unpaired and exactly equal to zero, with an eigenstate that is largely localized in the first Fock state. The results obtained in Sec.~\ref{Sec:Simulations}, and also in Ref.~\cite{Ashhab2026PRA}, fit this pattern. The increase in the magnitude of the matrix elements in $\hat{H}(n)$ is not as extreme as in $\hat{H}_{\rm hierarchical}$, especially as we go to higher Fock states. However, our numerical simulations show that for $n\geq 3$ the states at the center of the spectrum follow the general behaviour produced by the simple example of $\hat{H}_{\rm hierarchical}$. This toy model also provides an intuitive explanation for the seemingly paradoxical situation that occurs for $n>4$, where the dynamics is very sensitive to the parity of $N$, even though the photon number remains largely confined to small values, no matter how large $N$ is.

\subsection{Rotational symmetry of generalized squeezed states}

One might wonder what Q and Wigner functions would be obtained in the dynamics with fractional values of $n$. In particular, the phase-space functions have $n$-fold rotational symmetry. Examples of such rotationally symmetric functions are shown in Figs.~2 and 5 of Ref.~\cite{Ashhab2025} for the cases $n=3$ and $n=4$, respectively. These symmetries are a result of the fact that only multiples of $n$ photons appear in the quantum state. The symmetry in each pattern does not depend on the exact amplitudes of the different quantum states. Since we do not have a simple mapping between the basis states and photon number states in the case of fractional $n$, we cannot construct similar phase-space functions in a straightforward manner. Needless to say, it is not clear how one might visualize a fractional rotational symmetry, e.g.~a 3.71-fold rotational symmetry.

\section{Conclusion}
\label{Sec:Conclusion}

In conclusion, we have generalized calculations of generalized squeezing to include fractional values of the squeezing order $n$, which has allowed us to analyze some key features in the spectrum and dynamics associated with generalized squeezing. In particular, using fractional values of $n$ combined with extrapolation has allowed us to infer results about the computationally challenging cases $n=2$ and $n=4$, which are critical points at which qualitative changes in behaviour occur. In addition to its utility as a computational tool to predict the behaviour near such critical points, our approach has also allowed us to deduce intuitive explanations for phenomena related to the continuity of the spectrum and the oscillatory dynamics encountered in the study of generalized squeezing. These results add valuable insight to our understanding of high-order nonlinearities in quantum optics and suggest new methods to study related problems, such as the multi-photon quantum Rabi model.

\section*{Acknowledgments}

We would like to thank Mohammad Ayyash and Cid Reyes Bustos for useful discussions. This work was supported by Japan's Ministry of Education, Culture, Sports, Science and Technology (MEXT) Quantum Leap Flagship Program Grant Number JPMXS0120319794.

\section*{Data availability}

The datasets generated and/or analysed during the current study are available from the corresponding author on reasonable request.

\end{document}